\documentclass[10pt,a4papper]{article}
\usepackage{indentfirst}
\usepackage[]{graphicx}
\usepackage{geometry} % Меняем поля страницы
\begin{document}

\title{\textbf{The production of the $f_1(1285) \gamma$ and $a_1(1260) \gamma$ in colliding $e^{+}e^{-}$-beams in threshold domain}}
\author{M. K. Volkov\footnote{volkov@theor.jinr.ru}, A. A. Pivovarov\footnote{tex$\_$k@mail.ru} and A. A. Osipov\footnote{osipov@nu.jinr.ru}\\
\small
\emph{Bogoliubov Laboratory of Theoretical Physics, Joint Institute for Nuclear Research, Dubna, 141980, Russia}}
\date{}
\maketitle
\small

\begin{abstract}
The processes $e^{+}e^{-}\!\to\! (f_1(1285), a_1(1260))\gamma$ in threshold domain are considered in the framework of the extended Nambu--Jona-Lasinio model. The channels with the ground $\rho(770)$, $\omega(782)$ and radially excited $\rho(1450)$, $\omega(1420)$ intermediate meson states are taken into account. It is shown that in the process $e^{+}e^{-}\to f_1(1285)\gamma$, the probability of the subprocesses with $\rho$-mesons significantly exceeds the probability of the subprocesses with $\omega$-mesons, whereas, in the process $e^{+}e^{-} \to a_1(1260) \gamma$, $\rho$- and $\omega$-channels give approximately equal contributions. The mechanism of this effect is discussed. The radiative decay widths of $\rho(1450)\!\to\! f_{1}(1285)\gamma$, $\omega(1420)\!\to\! f_{1}(1285)\gamma$, $\rho(1450)\!\to\! a_1(1260)\gamma$ and $\omega(1420)\!\to\! a_1(1260)\gamma$ are calculated.
\end{abstract}
\large
\vspace{0.5cm}

\section{Introduction}
At present time, there are many both theoretical and experimental works devoted to an investigation of the production of pseudoscalar mesons with the photon in colliding $e^{+}e^{-}$-beams at low energies. As a rule, these processes can be described by the anomalous quark-loop vertices preceded by the single virtual photon or the hadron resonance state. At the energies below $2$ GeV it can be the $q\bar q$ resonance belonging to the ground $^3S_1$ vector-meson nonet, or its exited states. Because of inapplicability of the perturbative QCD at low energies 
it has become customary to describe the effects of resonance states within a phenomenological chiral Lagrangian framework. One of the most efficient models of this type is the Nambu--Jona-Lasinio (NJL) model \cite{Ebert:1982pk, Volkov:1984kq, Volkov:1986zb, Ebert:1985kz, Vogl:1991qt, Klevansky:1992qe, Hatsuda:1994pi, Ebert:1994mf}. The other alternative is the chiral perturbation theory \cite{Gasser84,Gasser85a,Gasser85b,Ecker89a,Ecker89b}.

Among the above mentioned well known processes of $e^{+}e^{-}$-annihilation into pseudoscalar meson and photon one can point out the following ones $e^{+}e^{-} \to (\pi^{0}, \pi(1300)) \gamma$ \cite{Dolinsky:1991vq, Achasov:1990at, Achasov:2003ed, Arbuzov:2011fv}, $e^{+}e^{-} \to (\eta, \eta^{'}(958), \eta(1295), \eta(1475)) \gamma$ \cite{Achasov:2006dv, Ahmadov:2013ksa}. Unfortunately, by the present time, the modes involving the axial-vector mesons have not been investigated well enough. To fill this gap, we study here the $e^+e^-$-production of the pairs $f_1(1285) \gamma$ and $a_1(1260) \gamma$ by using the extended Namby--Jona-Lasinio model \cite{Volkov:1996br, Volkov:1996fk, Volkov:2006vq, Volkov:2016umo}. These processes proceed through the anomalous quark loops. Such vertices have been considered recently in \cite{Osipov:2017ray}. We obtain here, additionally, the radiative decay widths of exited $\rho$- and $\omega$-like vector mesons $(\rho(1450), \omega(1420)) \to (f_1(1285), a_1(1260)) \gamma$. 

\section{The Lagrangian of the extended NJL model}
Let us consider the extended NJL model described by the effective quark-meson Lagrangian \cite{Volkov:2016umo}. For our purpose we only need the following part of it
\begin{eqnarray}
\label{Lagrangiane}
	\Delta L_{int} & = & \frac{1}{2}\,
						\bar{q} \left[\gamma^{\mu}\gamma^{5} \tau_{0} \left(A_{f_{1}}f_{1\mu} + B_{f_{1}}f^{'}_{1\mu}\right)
										+ \gamma^{\mu}\gamma^{5} \tau_{3} \left(A_{a_{1}}a_{1\mu} + B_{a_{1}}a^{'}_{1\mu}\right)\right. \nonumber \\
					&&  \left. + \gamma^{\mu} \tau_{3} \left(A_{\rho}\rho_{\mu} + B_{\rho}\rho^{'}_{\mu}\right)
						+ \gamma^{\mu} \tau_{0} \left(A_{\omega}\omega_{\mu} + B_{\omega}\omega^{'}_{\mu}\right)\right]q.
\end{eqnarray}
Here $q$ is the constituent quark field. Its $SU(2)$ flavor components are $u$- and $d$- constituent quarks with masses $m_{u} = m_{d} = 280$ MeV \cite{Volkov:1986zb}; $f_{1}$, $a_{1}$, $\rho$ and $\omega$ are the axial-vector and vector mesons; the corresponding excited states are marked with prime,
\begin{eqnarray}
A_{M} & = & \frac{1}{\sin(2\theta_{M}^{0})}\left[g_{M}\sin(\theta_{M} + \theta_{M}^{0}) + g_{M}^{'}f(\vec{k}^{2})\sin(\theta_{M} - \theta_{M}^{0})\right], \nonumber \\
B_{M} & = & \frac{-1}{\sin(2\theta_{M}^{0})}\left[g_{M}\cos(\theta_{M} + \theta_{M}^{0}) + g_{M}^{'}f(\vec{k}^{2})\cos(\theta_{M} - \theta_{M}^{0})\right],
\end{eqnarray}
$f\left(\vec{k}^{2}\right) = 1 + d \vec{k}^{2}$ is the form factor describing the first radially excited states \cite{Volkov:1996br, Volkov:1996fk}; $d$ is the slope parameter; $\theta_{M}$ and $\theta_{M}^{0}$ are the mixing angles of mesons in the ground and excited state \cite{Volkov:2016umo},
\begin{eqnarray}
	& \theta_{\rho} = \theta_{\omega} = \theta_{f_{1}} = \theta_{a_{1}} = 81.8^{\circ} & \nonumber\\
	& \theta_{\rho}^{0} = \theta_{\omega}^{0} = \theta_{f_{1}}^{0} = \theta_{a_{1}}^{0} = 61.5^{\circ} & \nonumber\\
	& d = -1.784 \textrm{GeV}^{-2}. &
\end{eqnarray}
The $U(2)$ matrices $\tau_3$ and $\tau_0$ are given by
\begin{eqnarray}
	\tau_{3} = \left(\begin{array}{cc}
	1 & 0  \\
	0 & -1 \\
	\end{array}\right), \quad
	\tau_{0} = \left(\begin{array}{cc}
	1 & 0  \\
	0 & 1  \\
	\end{array}\right).
\end{eqnarray}
The coupling constants are
\begin{eqnarray}
g_{\rho} = g_{\omega} = g_{f_{1}} = g_{a_{1}} = \left(\frac{2}{3}I_{2}\right)^{-1/2} \approx 6.14, \nonumber\\
g_{\rho}^{'} = g_{\omega}^{'} = g_{f_{1}}^{'} = g_{a_{1}}^{'} = \left(\frac{2}{3}I_{2}^{f^{2}}\right)^{-1/2} \approx 9.87,
\end{eqnarray}

\begin{equation}
I_{2}^{f^{n}} =
-i\frac{N_{c}}{(2\pi)^{4}}\int\frac{f^{n}(\vec{k}^{2})}{(m^{2} - k^2)^{2}}\theta(\Lambda_{3}^{2} - \vec{k}^2)
\mathrm{d}^{4}k,
\end{equation}
where $\Lambda_{3} = 1.03$ GeV \cite{Volkov:2016umo} is the cut off parameter.

\section{The radiative decays $(\rho(1450), \omega(1420)) \to (f_1(1285), a_1(1260)) \gamma$}
\begin{figure}[h]
	\centering\includegraphics[scale = 0.4]{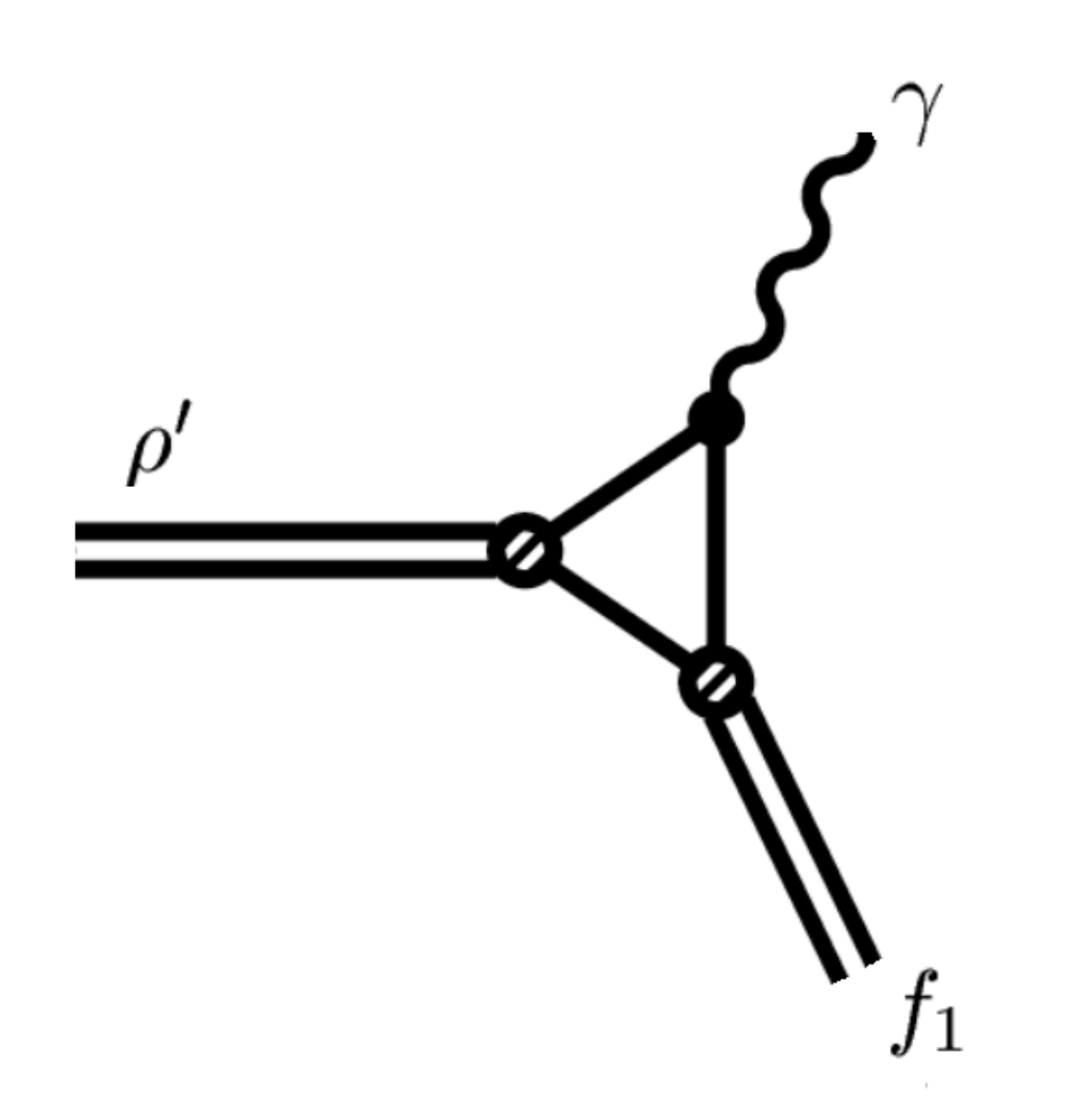}
	\caption{The decay  $\rho(1450) \to f_{1}(1285) \gamma$.}
	\label{Decay}
\end{figure}

By using the expressions for the appropriate vertices obtained in \cite{Osipov:2017ray} one can calculate the width of the radiative decay $\rho(1450) \to f_{1}(1285) \gamma$. The corresponding Feynman diagram is shown in Fig. \ref{Decay}. The amplitude is

\begin{eqnarray}
	\mathcal{M} = \frac{e}{2} e_{\mu}(p_{f_{1}}) e_{\nu}(p_{\rho}) e_{\lambda}(p_{\gamma}) I^{\mu\nu\lambda}_{(f_{1}\rho')},
\end{eqnarray}
where $e$ is the elementary charge, $\alpha_{em}=e^2/(4\pi)=1/137$; $e_{\mu}(p)$ is the polarisation vector of the spin-1 particle with 4-momentum $p$, $I^{\mu\nu\lambda}_{(f_{1}\rho')}$ is the integral which comes from the quark triangle calculations
\begin{eqnarray}
I^{\mu\nu\lambda}_{(f_{1}\rho')} & = & \frac{4}{3} \left\{-e^{\mu\nu\lambda p_{\gamma}} \left[2p_{\rho}^{2} - \left(p_{\rho}, p_{\gamma}\right)\right] + e^{\mu\nu\lambda p_{\rho}} \left(p_{\rho}, p_{\gamma}\right) - e^{\mu\nu p_{\gamma} p_{\rho}} p_{\rho}^{\lambda} - e^{\lambda\mu  p_{\rho} p_{\gamma}} p_{\gamma}^{\nu}\right\} \nonumber\\
&& \times \left\{-I_{3}^{(f_{1}\rho')} + 2m^{2}I_{4}^{(f_{1}\rho')} - m^{4}I_{5}^{(f_{1}\rho')}\right\},
\end{eqnarray}
where $p_{\rho}$ and $p_{\gamma}$ are the momenta of the $\rho$-meson and the photon, the scalar parts are given by 
\begin{equation}
I_{n}^{(f_{1}\rho')} = -i\frac{N_{c}}{(2\pi)^{4}}\int\frac{A_{f_{1}} B_{\rho}}{(m^{2} - k^2)^{n}}\theta(\Lambda_{3}^{2} - \vec{k}^2) \mathrm{d}^{4}k,
\end{equation}
where $a_{f_{1}}$ and $b_{\rho}$ are the couplings of the Lagrangian density (\ref{Lagrangiane}).

The decay width of the process takes the form
\begin{eqnarray}
\Gamma = \frac{\alpha_{em}}{54} \frac{(M_{\rho'}^{2} + M_{f_{1}}^{2}) (M_{\rho'}^{2} - M_{f_{1}}^{2})^{3}}{M_{\rho'} M_{f_{1}}^{2}}   \left[-I_{3}^{(f_{1}\rho')} + 2m^{2}I_{4}^{(f_{1}\rho')} - m^{4}I_{5}^{(f_{1}\rho')}\right]^{2}.
\end{eqnarray}

Similarly, one can obtain the decay widths of the processes $\omega(1420)\to f_{1}(1285) \gamma$, $\rho(1450) \to a_{1}(1260) \gamma$ and $\omega(1420) \to a_{1}(1260) \gamma$. The results are given in the Table \ref{Tab}.\\

\begin{table}[h]
\caption{The decay widths of the processes $(\rho(1450), \omega(1420)) \to (f_{1}(1285), a_{1}(1260)) \gamma$.}
\label{Tab}
\begin{center}
\begin{tabular}{|c|c|c|}
\hline
& $f_{1}(1285) \gamma$ 	& $a_{1}(1260) \gamma$  \\
\hline
$\rho(1450) \to$   & 1.43 keV & 0.33 keV \\
$\omega(1420) \to$ & 0.07 keV &	1.63 keV \\
\hline
\end{tabular}
\end{center}
\end{table}

\section{The processes $e^{+}e^{-} \to (f_1(1285), a_1(1260))\gamma$}
\begin{figure}[h]
\centering\includegraphics[scale = 0.5]{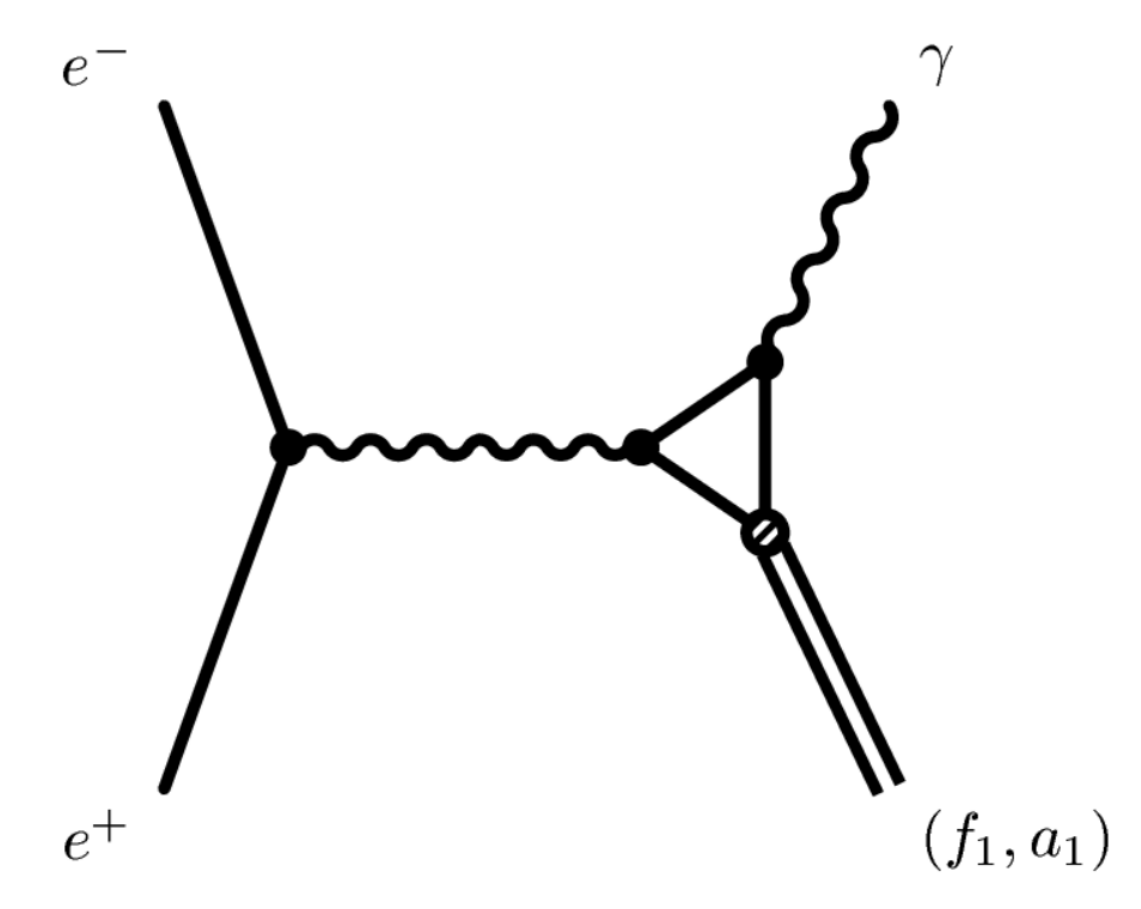}
\caption{The Feynman ``contact'' diagrams for the reactions $e^{+}e^{-} \to (f_1(1285), a_1(1260))\gamma$ with the intermediate photon.}
\label{Contact}
\end{figure}

\begin{figure}[h]
\centering\includegraphics[scale = 0.7]{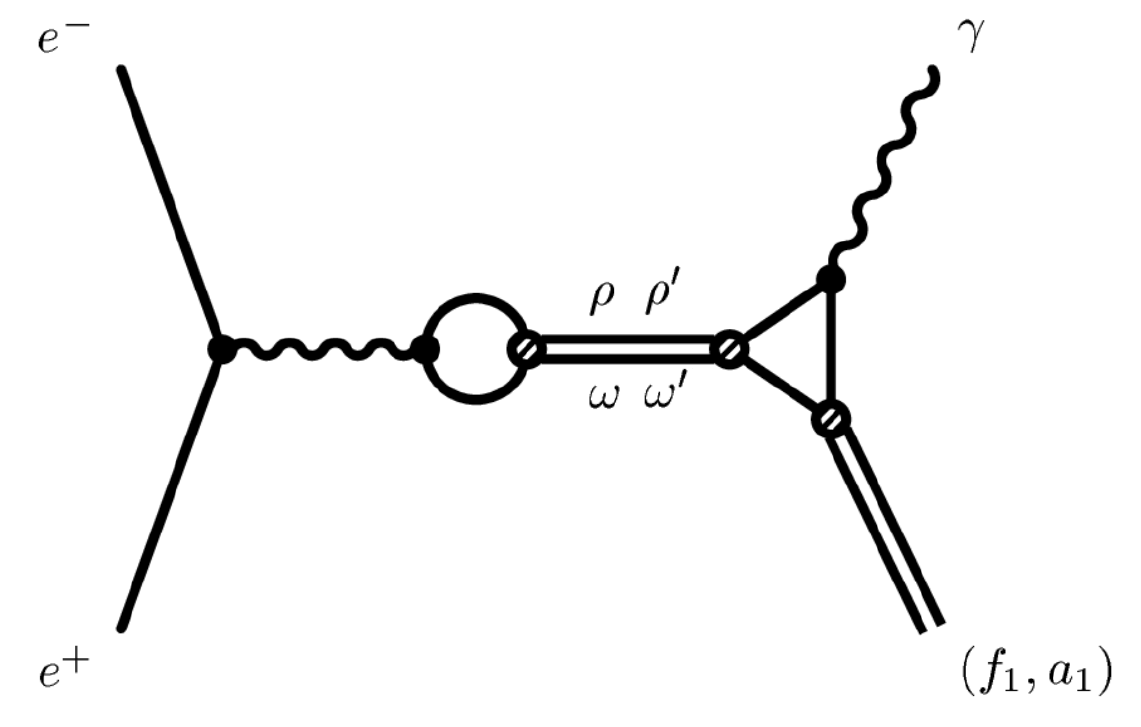}
\caption{The Feynman diagrams with the intermediate vector mesons $\rho(770)$, $\rho(1450)$, $\omega(782)$ and $\omega(1420)$ contributing to the amplitudes of the reactions $e^{+}e^{-} \to (f_1(1285), a_1(1260))\gamma$.}
\label{Intermediate}
\end{figure}

The diagrams of the processes $e^{+}e^{-} \to (f_1(1285), a_1(1260))\gamma$ are shown in Figs. \ref{Contact} and \ref{Intermediate}. The corresponding amplitudes are


\begin{eqnarray}
\label{amplitudes1}
&&\mathcal{M}(e^{+}e^{-} \to f_1(1285)\gamma) = \frac{e^{3}}{s} l_{\nu} \left\{\frac{5}{9} I_{(f_{1})}^{\mu\nu\lambda} + \frac{C_{\rho}}{2g_{\rho}} \frac{s}{M_{\rho}^{2} - s - i\sqrt{s}\Gamma_{\rho}}I_{(f_{1}\rho)}^{\mu\nu\lambda} + \frac{C_{\rho'}}{2g_{\rho}} \frac{s}{M_{\rho'}^{2} - s - i\sqrt{s}\Gamma_{\rho'}}I_{(f_{1}\rho')}^{\mu\nu\lambda}\right. \nonumber \\
&& \left.+ \frac{C_{\omega}}{18g_{\omega}} \frac{s}{M_{\omega}^{2} - s - i\sqrt{s}\Gamma_{\omega}}I_{(f_{1}\omega)}^{\mu\nu\lambda} + \frac{C_{\omega'}}{18g_{\omega}} \frac{s}{M_{\omega'}^{2} - s - i\sqrt{s}\Gamma_{\omega'}}I_{(f_{1}\omega')}^{\mu\nu\lambda}\right\} e(p_{f_{1}})_{\mu} e(p_{\gamma})_{\lambda}, \nonumber \\	
&&\mathcal{M}(e^{+}e^{-} \to a_1(1260)\gamma) = \frac{e^{3}}{s} l_{\nu} \left\{\frac{1}{3} I_{(a_{1})}^{\mu\nu\lambda} + \frac{C_{\rho}}{6g_{\rho}} \frac{s}{M_{\rho}^{2} - s - i\sqrt{s}\Gamma_{\rho}}I_{(a_{1}\rho)}^{\mu\nu\lambda} + \frac{C_{\rho'}}{6g_{\rho}} \frac{s}{M_{\rho'}^{2} - s - i\sqrt{s}\Gamma_{\rho'}}I_{(a_{1}\rho')}^{\mu\nu\lambda}\right. \nonumber \\
&&  \left.+ \frac{C_{\omega}}{6g_{\omega}} \frac{s}{M_{\omega}^{2} - s - i\sqrt{s}\Gamma_{\omega}}I_{(a_{1}\omega)}^{\mu\nu\lambda} + \frac{C_{\omega'}}{6g_{\omega}} \frac{s}{M_{\omega'}^{2} - s - i\sqrt{s}\Gamma_{\omega'}}I_{(a_{1}\omega')}^{\mu\nu\lambda}\right\} e(p_{a_{1}})_{\mu} e(p_{\gamma})_{\lambda},
\end{eqnarray}
where $l^{\mu} = \bar{e}\gamma^{\mu}e$ is the lepton current, $M_{\rho} = 775$ MeV, $M_{\rho^{'}} = 1465$ MeV, $M_{\omega} = 783$ MeV, $M_{\omega^{'}} = 1425$ MeV, $\Gamma_{\rho} = 148$ MeV, $\Gamma_{\rho^{'}} = 400$ MeV, $\Gamma_{\omega} = 8$ MeV, $\Gamma_{\omega^{'}} = 215$ MeV are the masses and the full widths of the intermediate vector mesons \cite{Agashe:2014kda},
\begin{eqnarray}
C_{\rho} = C_{\omega} & = & \frac{1}{\sin\left(2\theta_{\rho}^{0}\right)}\left[\sin\left(\theta_{\rho} + \theta_{\rho}^{0}\right) +
				R_{V}\sin\left(\theta_{\rho} - \theta_{\rho}^{0}\right)\right], \nonumber \\
C_{\rho^{'}} = C_{\omega^{'}} & = & \frac{-1}{\sin\left(2\theta_{\rho}^{0}\right)}\left[\cos\left(\theta_{\rho} + \theta_{\rho}^{0}\right) +
				R_{V}\cos\left(\theta_{\rho} - \theta_{\rho}^{0}\right)\right], \nonumber \\
R_{V}	 & = & \frac{I_{2}^{f}}{\sqrt{I_{2}I_{2}^{f^{2}}}}.
\end{eqnarray}

In (\ref{amplitudes1}), the first terms correspond to the contact diagrams, the other four terms describe the contributions of the intermediate $\rho(770)$, $\rho(1450)$, $\omega(782)$ and $\omega(1420)$ mesons. It is useful to divide these amplitudes on the $\rho$- and $\omega$-channels by combining of the appropriate parts of the contact diagram with the other contributions:
\begin{eqnarray}
\mathcal{M}(e^{+}e^{-} \to f_1(1285)\gamma) &\! =\! & \frac{e^{3}}{s} l_{\nu} \left\{M_{f_{1}\rho}^{\mu\nu\lambda} + M_{f_{1}\omega}^{\mu\nu\lambda}\right\}  e(p_{f_{1}})_{\mu} e(p_{\gamma})_{\lambda}, \nonumber\\
\mathcal{M}(e^{+}e^{-} \to a_1(1260)\gamma) &\! =\! & \frac{e^{3}}{s} l_{\nu} \left\{M_{a_{1}\rho}^{\mu\nu\lambda} + M_{a_{1}\omega}^{\mu\nu\lambda}\right\}  e(p_{a_{1}})_{\mu} e(p_{\gamma})_{\lambda},
\end{eqnarray}
where
\begin{eqnarray}
M_{f_{1}\rho}^{\mu\nu\lambda} &\! =\! & \frac{1}{2} \left\{I_{(f_{1})}^{\mu\nu\lambda}
+ \frac{C_{\rho}}{g_{\rho}} \frac{s}{M_{\rho}^{2} - s - i\sqrt{s}\Gamma_{\rho}}I_{(f_{1}\rho)}^{\mu\nu\lambda} + \frac{C_{\rho'}}{g_{\rho}} \frac{s}{M_{\rho'}^{2} - s - i\sqrt{s}\Gamma_{\rho'}}I_{(f_{1}\rho')}^{\mu\nu\lambda}\right\}, \nonumber\\
M_{f_{1}\omega}^{\mu\nu\lambda} &\! =\!& \frac{1}{18} \left\{I_{(f_{1})}^{\mu\nu\lambda}
+ \frac{C_{\omega}}{g_{\omega}} \frac{s}{M_{\omega}^{2} - s - i\sqrt{s}\Gamma_{\omega}}I_{(f_{1}\omega)}^{\mu\nu\lambda} + \frac{C_{\omega'}}{g_{\omega}} \frac{s}{M_{\omega'}^{2} - s - i\sqrt{s}\Gamma_{\omega'}}I_{(f_{1}\omega')}^{\mu\nu\lambda}\right\}, \nonumber\\
M_{a_{1}\rho}^{\mu\nu\lambda} &\! =\! & \frac{1}{6} \left\{I_{(a_{1})}^{\mu\nu\lambda}
+ \frac{C_{\rho}}{g_{\rho}} \frac{s}{M_{\rho}^{2} - s - i\sqrt{s}\Gamma_{\rho}}I_{(a_{1}\rho)}^{\mu\nu\lambda} + \frac{C_{\rho'}}{g_{\rho}} \frac{s}{M_{\rho'}^{2} - s - i\sqrt{s}\Gamma_{\rho'}}I_{(a_{1}\rho')}^{\mu\nu\lambda}\right\}, \nonumber\\
M_{a_{1}\omega}^{\mu\nu\lambda} &\! =\! & \frac{1}{6} \left\{I_{(a_{1})}^{\mu\nu\lambda}
+ \frac{C_{\omega}}{g_{\omega}} \frac{s}{M_{\omega}^{2} - s - i\sqrt{s}\Gamma_{\omega}}I_{(a_{1}\omega)}^{\mu\nu\lambda} + \frac{C_{\omega'}}{g_{\omega}} \frac{s}{M_{\omega'}^{2} - s - i\sqrt{s}\Gamma_{\omega'}}I_{(a_{1}\omega')}^{\mu\nu\lambda}\right\}.
\end{eqnarray}

By using these amplitudes one can obtain the cross sections of the considered processes:
\begin{eqnarray}
\sigma(e^{+}e^{-} \to f_1(1285)\gamma) & = & \frac{32}{27}\pi^{2}\alpha_{em}^{3} \frac{(s - M_{f_{1}}^{2})^{3}(s + M_{f_{1}}^{2})}{s^{2} M_{f_{1}}^{2}}|S_{f_{1}\rho} + S_{f_{1}\omega}|^{2}, \nonumber\\
\sigma(e^{+}e^{-} \to a_1(1260)\gamma) & = & \frac{32}{27}\pi^{2}\alpha_{em}^{3} \frac{(s - M_{a_{1}}^{2})^{3}(s + M_{a_{1}}^{2})}{s^{2} M_{a_{1}}^{2}} |S_{a_{1}\rho} + S_{a_{1}\omega}|^{2},
\end{eqnarray}
where $S_{f_{1}\rho}$, $S_{f_{1}\omega}$, $S_{a_{1}\rho}$ and $S_{a_{1}\omega}$ are the scalar parts of the appropriate amplitudes:
\begin{eqnarray}
S_{f_{1}\rho} 	& = & \frac{1}{2} \left\{\left[-I_{3}^{(f_{1})} + 2m^{2}I_{4}^{(f_{1})} - m^{4}I_{5}^{(f_{1})}\right] + \frac{C_{\rho}}{g_{\rho}} \frac{s}{M_{\rho}^{2} - s - i\sqrt{s}\Gamma_{\rho}}\left[-I_{3}^{(f_{1}\rho)} + 2m^{2}I_{4}^{(f_{1}\rho)} - m^{4}I_{5}^{(f_{1}\rho)}\right]\right. \nonumber\\
&& \left. + \frac{C_{\rho'}}{g_{\rho}} \frac{s}{M_{\rho'}^{2} - s - i\sqrt{s}\Gamma_{\rho'}}\left[-I_{3}^{(f_{1}\rho')} + 2m^{2}I_{4}^{(f_{1}\rho')} - m^{4}I_{5}^{(f_{1}\rho')}\right]\right\}, \nonumber\\
S_{f_{1}\omega} & = & \frac{1}{18} \left\{\left[-I_{3}^{(f_{1})} + 2m^{2}I_{4}^{(f_{1})} - m^{4}I_{5}^{(f_{1})}\right] + \frac{C_{\omega}}{g_{\omega}} \frac{s}{M_{\omega}^{2} - s - i\sqrt{s}\Gamma_{\omega}}\left[-I_{3}^{(f_{1}\omega)} + 2m^{2}I_{4}^{(f_{1}\omega)} - 	m^{4}I_{5}^{(f_{1}\omega)}\right]\right. \nonumber\\
&& \left.+ \frac{C_{\omega'}}{g_{\omega}} \frac{s}{M_{\omega'}^{2} - s - i\sqrt{s}\Gamma_{\omega'}}\left[-I_{3}^{(f_{1}\omega')} + 2m^{2}I_{4}^{(f_{1}\omega')} - 	m^{4}I_{5}^{(f_{1}\omega')}\right]\right\}, \nonumber\\
S_{a_{1}\rho}& = & \frac{1}{6} \left\{\left[-I_{3}^{(a_{1})} + 2m^{2}I_{4}^{(a_{1})} - m^{4}I_{5}^{(a_{1})}\right] + \frac{C_{\rho}}{g_{\rho}} \frac{s}{M_{\rho}^{2} - s - i\sqrt{s}\Gamma_{\rho}}\left[-I_{3}^{(a_{1}\rho)} + 2m^{2}I_{4}^{(a_{1}\rho)} - m^{4}I_{5}^{(a_{1}\rho)}\right]\right. \nonumber\\
&& \left.+ \frac{C_{\rho'}}{g_{\rho}} \frac{s}{M_{\rho'}^{2} - s - i\sqrt{s}\Gamma_{\rho'}}\left[-I_{3}^{(a_{1}\rho')} + 2m^{2}I_{4}^{(a_{1}\rho')} - 	m^{4}I_{5}^{(a_{1}\rho')}\right]\right\}, \nonumber\\
S_{a_{1}\omega}	& = & \frac{1}{6} \left\{\left[-I_{3}^{(a_{1})} + 2m^{2}I_{4}^{(a_{1})} - m^{4}I_{5}^{(a_{1})}\right] + \frac{C_{\omega}}{g_{\omega}} \frac{s}{M_{\omega}^{2} - s - i\sqrt{s}\Gamma_{\omega}}\left[-I_{3}^{(a_{1}\omega)} + 2m^{2}I_{4}^{(a_{1}\omega)} - 	m^{4}I_{5}^{(a_{1}\omega)}\right]\right. \nonumber\\
&& \left. + \frac{C_{\omega'}}{g_{\omega}} \frac{s}{M_{\omega'}^{2} - s - i\sqrt{s}\Gamma_{\omega'}}\left[-I_{3}^{(a_{1}\omega')} + 2m^{2}I_{4}^{(a_{1}\omega')} - 	m^{4}I_{5}^{(a_{1}\omega')}\right]\right\}.
\end{eqnarray}

The values of the cross sections for the processes $e^{+}e^{-}\!\to\! (f_1(1285), a_1(1260))\gamma$ for some values of $\sqrt{s}$ are pointed in the Table~\ref{Tab2}.\\

\begin{table}[h]
	\caption{The values of the cross sections for the processes $e^{+}e^{-}\!\to\! (f_1(1285), a_1(1260))\gamma$.}
	\label{Tab2}
	\begin{center}
		\begin{tabular}{|c|c|c|c|}
			\hline
			$\sqrt{s}$, MeV & 1350 & 1450 & 1550 \\
			\hline
			$\sigma(e^{+}e^{-}\!\to\! f_1(1285) \gamma)$, pb & 0.068 & 0.5 & 0.901 \\
			$\sigma(e^{+}e^{-}\!\to\! a_1(1260) \gamma)$, pb & 0.15 & 0.398 & 0.474 \\
			\hline
		\end{tabular}
	\end{center}
\end{table}

\begin{figure}[h]
\centering\includegraphics[scale = 0.5]{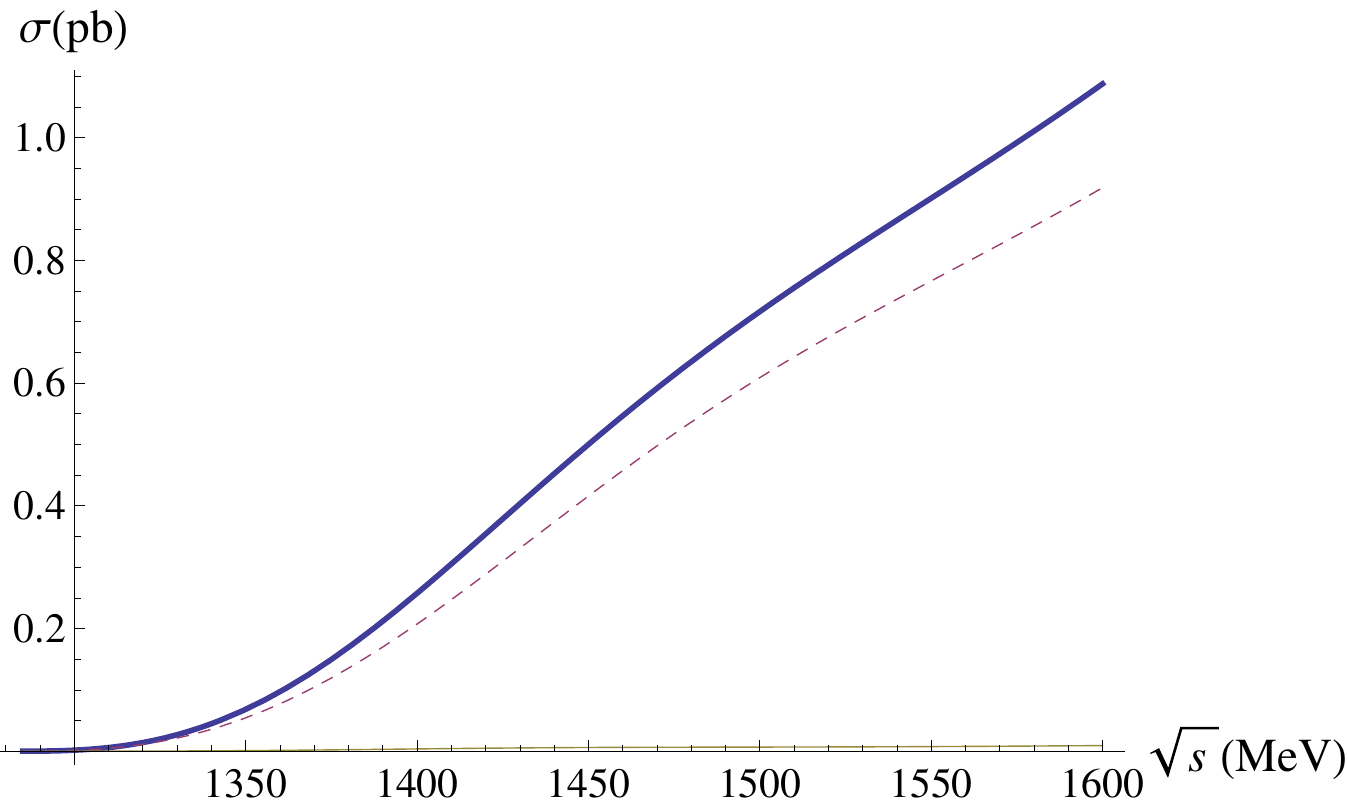}
\caption{The cross section of the process $e^{+}e^{-} \to f_1(1285)\gamma$. The thick line corresponds to the total cross section, the dashed line corresponds to the $\rho$-channel (diagram with intermediate $\rho$-mesons + appropriate part of the contact diagram), the thin line corresponds to the $\omega$-channel (diagram with intermediate $\omega$-mesons + appropriate part of the contact diagram).}
\label{CrossSection1}
\end{figure}

\begin{figure}[h]
\centering\includegraphics[scale = 0.5]{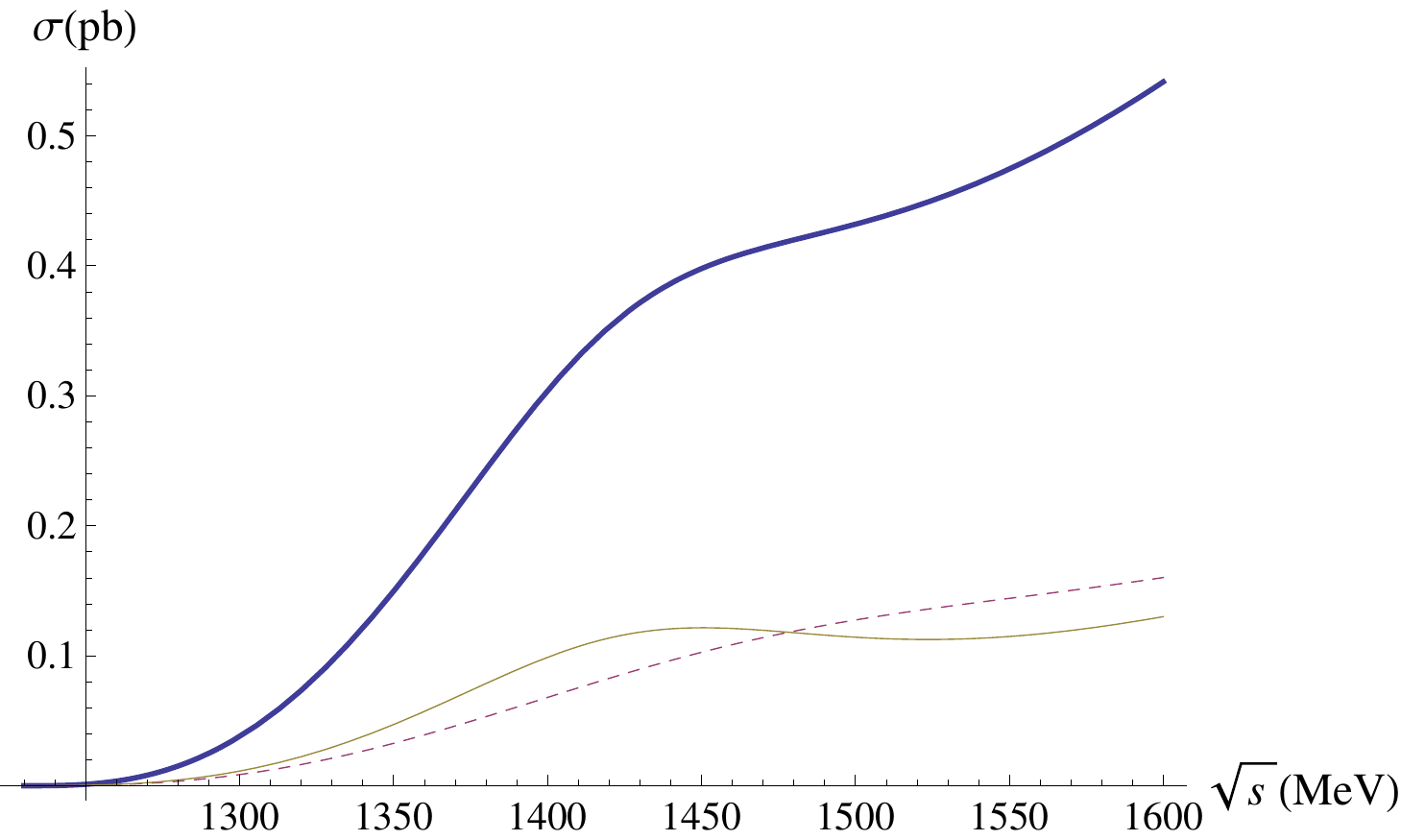}
\caption{The cross section of the process $e^{+}e^{-} \to a_1(1260)\gamma$. The thick line corresponds to the total cross section, the dashed line corresponds to the $\rho$-channel (diagram with intermediate $\rho$-mesons + appropriate part of the contact diagram), the thin line corresponds to the $\omega$-channel (diagram with intermediate $\omega$-mesons + appropriate part of the contact diagram).}
\label{CrossSection2}
\end{figure}

The dependencies of the cross section on the energy of colliding leptons, $\sqrt{s}$, for the processes $e^{+}e^{-} \to (f_1(1285), a_1(1260))\gamma$ are shown in Figs. \ref{CrossSection1} and \ref{CrossSection2}. The dashed lines correspond to the $\rho$-channel, the thin lines correspond to the $\omega$-channel. The thick lines show the total contribution. In Fig. \ref{CrossSection1}, the $\omega$-channel is two orders lower and almost not seen.


Our calculations show that the contributions from the subprocesses with $\rho$- and $\omega$-mesons to the process $e^{+}e^{-} \to a_1(1260)\gamma$ are of the same order. Whereas, in the process $e^{+}e^{-} \to f_1(1285)\gamma$, the contribution from the subprocess with $\rho$-mesons significantly exceeds the contribution of $\omega$-mesons. This behaviour may be explained as follows. If we proceed from Vector Dominance Model, the photon in the final state may be considered as produced from the vector meson. In the process $e^{+}e^{-} \to a_1(1260)\gamma$, the productions of $a_1(1260)\gamma$ from $\rho$- and $\omega$- intermediate mesons are described by the same vertex $a_{1}^{0}\rho^{0}\omega$ where $\rho$- or $\omega$-mesons turn into the photon. In the process $e^{+}e^{-} \to f_1(1285)\gamma$, we have two different vertices which describe transition of the intermediate vector meson into $f_{1}(1285)\gamma$. Indeed, in the case of $\rho$-channel, the vertex $f_{1}\rho^{0}\rho^{0}$ is used where one of $\rho$-mesons turn into the final photon. In the case of $\omega$-channel, the process contains the vertex $f_{1}\omega\omega$ where one of $\omega$-mesons goes into the final photon. Each $\omega\to\gamma$ transition contains the factor $1/3$ \cite{Volkov:1986zb}. We can see that in the process $e^{+}e^{-} \to f_1(1285)\gamma$, in the $\omega$-channel, this factor appears twice: in the vertex $\gamma \to \omega$ and in the vertex $\omega \to f_{1}\gamma$. As a result, in the $\omega$-channel, the factor $1/9$ appears in comparison with the $\rho$-channel. On the other hand, in the process $e^{+}e^{-} \to a_1(1260)\gamma$, the factor $1/3$ appears only once: in the vertex $\gamma \to \omega$ or in the vertex $\rho \to f_{1}\gamma$. Therefore, these channels make the same contributions.

\section{Discussion and Conclusion}
Due to absences the appropriate experimental data on the reactions discussed, it would be interesting to check our theoretical predictions in the future experiments on the modern accelerators such as VEPP-2000 (Novosibirsk), BEPC-II (Beijing), Belle (KEK, Japan), etc. Such experiments could shed light on the anomalous nature of the hadronic interactions studied in this paper. Unfortunately, the description presented here is valid only in a very restricted range of energies and, thus, should be corrected at energies $\sqrt{s}\geq 1.5$ GeV, where the new excited states $\rho(1570)$, $\omega(1650)$ must be included. We also did not take into account the mixing effects of the $f_1(1285)$, $f_1(1420)$ and $f_1(1510)$ mesons \cite{Dickson93}. The latter could not exert a strong influence on the results. Nevertheless, it would be interesting to take this mixing into account as soon as the first experimental data will be obtained.

\end{document}